\documentclass[aps,prl,reprint,amsmath,amssymb]{revtex4-1}

\usepackage{epsfig,color,graphicx}
\usepackage{algorithm}
\usepackage{algpseudocode}
\usepackage{soul, amsmath, amssymb} %, subfig, dblfloatfix}
\setcitestyle{super}
\usepackage{xr}
\usepackage{setspace}
\usepackage{multirow}
\usepackage{footnote}
\usepackage{scrextend}
\makeatletter
\renewcommand\NAT@biblabelnum[1]{{(#1)}}
\makeatother

\begin{document}

\nocite{achemso-ctrl}

\bibliographystyle{achemso} 

\title{Direct unconstrained optimization of excited states in density functional theory}
\author{Hanh D. M. Pham}
%\author{Zhenzhe Zhang} - providing optimized geometries would not qualify Zhenzhe for co-authorship, acknoledgement will be sufficient
\author{Rustam Z. Khaliullin}
%\email{rustam.khaliullin@mcgill.ca}
\affiliation{Department of Chemistry, McGill University, 801 Sherbrooke St West, Montreal, QC H3A 0B8, Canada}

\author{Email: rustam.khaliullin@mcgill.ca}

\date{\today}

%RZK: After thesis submission, prepare the standard supplementary data - optimized geometries and periodic cells.
%CINDY: DONE

\label{sec:abstract}

\begin{abstract}
\textbf{Abstract.} Orbital-optimized density functional theory (DFT) has emerged as an alternative to time-dependent (TD) DFT capable of describing difficult excited states with significant electron density redistribution, such as charge-transfer, Rydberg, and double-electron excitations. Here, a simple method is developed to solve the main problem of the excited-state optimization -- the variational collapse of the excited states onto the ground state. In this method, called variable-metric time-independent DFT (VM TIDFT), the electronic states are allowed to be nonorthogonal during the optimization but their orthogonality is gradually enforced with a continuous penalty function. With nonorthogonal electronic states, VM TIDFT can use molecular orbital coefficients as independent variables, which results in a closed-form analytical expression for the gradient and allows to employ any of the multiple unconstrained optimization algorithms that guarantees convergence of the excited-state optimization. Numerical tests on multiple molecular systems show that the variable-metric optimization of excited states performed with a preconditioned conjugate gradient algorithm is robust and produces accurate energies for well-behaved excitations and, unlike TDDFT, for more challenging charge-transfer and double-electron excitations.
%Furthermore, variational oribtal optimization makes it easy to compute the atomic forces in excited states, facilitating description of dynamical processes in excited states.
%The main advantages of VM TIDFT are its conceptual simplicity, utilization of the direct optimization that (unlike iterative eigensolver methods) can guarantee convergence, ability to optimize multiple excited states, the absence of convoluted parameterization of unitary transformations, simple closed-form expressions for the gradient, ability to perform full optimization of spin-pure open-shell singlet exicted states within the ROKS formalism, easy to compute the atomic forces in excited states.
\end{abstract}

\maketitle

\section{Introduction}

Excited electronic states play a crucial role in photophysics and photochemistry~\cite{matsika2017introduction, andres2005quantum, schlag1971lifetime}.
Theoretical and computational characterization of excited states is essential, not only from a fundamental science perspective but also for designing better photovoltaic, photocatalytic, and lighting materials.
One of the most popular methods to describe excited states is time-dependent (TD) density functional theory (TDDFT), which is an extension of the ground-state Kohn-Sham (KS) density functional theory (DFT). 
TDDFT calculates excited state properties by analyzing the response of the ground-state density to time-dependent perturbations, such as an external electric field~\cite{dreuw2005single}. 
Although TDDFT is, in principle, exact~\cite{runge1984density}, it has multiple limitations that arise from its reliance on time-independent ground state XC functionals~\cite{dreuw2005single}. 
Consequently, TDDFT cannot describe double- and higher-electron excitations and struggles with excitations producing significant changes of the electron density distribution, such as long-range charge transfer states, Rydberg states, and core excitations~\cite{hait2021orbital, ullrich2011time}.

To address the limitations of TDDFT, a variety of orbital optimized (OO) DFT methods have been developed. In these methods, the ground-state DFT formalism is applied to optimize orbitals variationally not only for the ground state but also for the excited states. To contrast this approach to TDDFT, it has been referred to as time-independent DFT (TIDFT)~\cite{levy1999variational, nagy2001variational, ayers2009time, ayers2012time}. 
$\Delta$SCF is another commonly used name for this approach~\cite{cullen2011formualtion, ziegler1977calculation,barth1979local, gavnholt2008delta, kowalczyk2011assessment, yang2024foundation} which emphasizes that the excitation energies are computed as the energy difference between self-consistent-field (SCF) calculations.
The variational treatment of the excited states yields excellent results, comparable to those obtained with TDDFT for well-behaved excited states and superior for charge-transfer (CT) and multi-electron excitations~\cite{triguero1999separate,cheng2008rydberg,seidu2015applications,liu2017evaluating, hait2020highly, kumar2022robust, selenius2024orbital}. 
Another advantage of using the ground-state formalism for excited states is that numerous techniques developed over several decades for KS DFT can be applied to excited states. For example, the implicit solvation models employed in ground states calculations can be readily used for excited states~\cite{kunze2021pcm, nottoli2023deltascf}.

While the effort to establish a solid theoretical foundation for the excited-state OODFT methods is underway~\cite{levy1999variational, nagy2001variational, ayers2009time, ayers2012time, yang2024foundation}, the most significant practical challenge in designing efficient OODFT methods is ensuring that the variational optimization converges to the correct excited state rather than collapsing back to the ground state~\cite{selenius2024orbital}. 
There has been a number of methods proposed to avoid variational collapse of high-energy excited states to the states with lower energies~\cite{gross1988density, gilbert2008self, baruah2009dft, kaduk2012constrained, ziegler2012implementation, evangelista2013orthogonality, filatov2015spin, ramos2018low, levi2020variational, levi2020variationalc, zhao2020density, fenk2020state, hait2020excited, roychoudhury2020neutral, ivanov2021method, macEtti2021initial, zhang2023target, schmerwitz2023calculations}. 
These methods have been recently reviewed~\cite{kaduk2012constrained, ziegler2015constricted, hait2021orbital, cernatic2022ensemble} and only a brief description of some of them is presented here.

% maximum overlap method (MOM) 2008
In the maximum overlap method (MOM)~\cite{gilbert2008self} and initial maximum overlap method (IMOM)~\cite{barca2018simple, barca2018excitation}, the electrons are first promoted from the fully optimized ground-state occupied MOs to virtual MOs to generate the initial guess for an excited state. The newly selected occupied MOs are then optimized in the SCF procedure, in each step of which the list of occupied orbitals is updated using a non-aufbau population principle.
In IMOM, the occupied orbitals are updated to maximize their overlap with the initial guess, whereas, in MOM, the orbitals are updated to produce maximum overlap with the orbitals in the previous SCF step. 
By maximizing the overlap with high-lying orbitals, MOM and IMOM prevent variational collapse to the ground state, guiding the SCF process toward the nearest excited state solution. Although both methods work well for multiple systems, MOM cannot always prevent variational collapse, as orbitals can drift to the ground state over multiple steps~\cite{mewes2014on, barca2018simple, barca2018excitation}. While this issue is addressed in IMOM, the latter exhibits its own convergence problems when multiple orbital selections have similar overlaps, which produces oscillatory non-convergent behavior~\cite{hait2020excited, fenk2020state}. 

% Orthogonality constrained OCDFT 2013
An alternative approach to prevent variational collapse is to constrain the KS determinant of an excited state to be orthogonal to that of the ground state. In practice, orthogonality-constrained DFT  (OCDFT)~\cite{evangelista2013orthogonality} recasts the optimization problem as a solution to modified eigenvector equations and imposes this constraint by projecting out a single ``hole'' orbital from the KS Hamiltonian. 
Unfortunately, the iterative solution of eigenvector problems is not guaranteed to convergence. Furthermore, the projector-based formalism of OCDFT makes its extension to multiple excited states cumbersome.

To improve the rate of convergence, several direct minimization algorithms have been developed. For example, the ground-state geometric direct minimization (GDM) approach~\cite{voorhis2002geometric} has been extended to optimize excited states with the IMOM orbital-selection criterion applied on each GDM iteration to prevent the collapse~\cite{levi2020variational, levi2020variationalc, ivanov2021method, schmerwitz2023calculations}. While the combination of GDM and IMOM resolves some of the convergence issues of the original eigensolver-based MOM, the combined method still does not guarantee convergence because the application of the IMOM criterion is introduced as an \emph{ad hoc} update, which can interfere with the GDM optimizer.
Another direct optimization approach is square gradient minimization (SGM)~\cite{hait2020excited}. SGM minimizes the square of the Frobenius norm of the energy gradient instead of minimizing the energy and thereby converts the saddle-point excited-state optimization into the minimization problem, obviating the need in any collapse prevention measures. Despite being superior to MOM-based methods in terms of convergence for difficult problems, SGM has its own shortcomings, as the norm of the gradient is less well conditioned than the energy, typically requiring more iterations to converge for simple cases~\cite{hait2020excited, shea2020generalized}. Furthermore, SGM can become trapped in local minima that are not stationary points of the energy functional~\cite{hait2020excited}.

% Introduce our work
This work presents a new approach to the variational optimization of excited electronic states, called variable-metric (VM) TIDFT. Like OCDFT, VM TIDFT prevents the variational collapse by imposing orthogonality constraints between electronic states. Unlike OCDFT, VM TIDFT allows nonorthogonal electronic states in the optimization process but pushes them towards orthogonality with a single continuous penalty function. The key advantage of VM TIDFT is the simplicity of its formalism that allows to relax multiple excited states using molecular orbital coefficients as independent variables in a direct unconstrained minimization that guarantees convergence of the SCF procedure. VM optimization has been designed specifically to prevent collapse of non-orthogonal wavefunctions and has been previously applied to improve localization of occupied and virtual orbitals~\cite{luo2020direct, luo2021variable} and to simplify the ground-state SCF procedure~\cite{pham2024direct}.

\section{Methodology}

\label{subsec:theory}

\textbf{Theory.} In our notation, capital Latin letters ($I, J, K, \ldots$) denote electronic states, lowercase Latin letters ($i, j, k, \ldots$) denote occupied molecular orbitals, and lowercase Greek letters ($\mu, \nu, \gamma, \lambda, \ldots$) represent basis set functions. Additionally, Greek letters $\alpha$ and $\beta$ are reserved for spin-up and spin-down electrons, respectively, whereas the symbol $\tau$ denotes a general spin state.

The unrestricted VM TIDFT formalism presented here optimizes $\alpha$ and $\beta$ orbitals independently to take into account the open-shell nature of excited states. However, restricted closed-shell (RKS), restricted-open shell (ROSO), and spin-purified restricted open-shell Kohn-Sham (ROKS)~\cite{frank1998molecular, filatov1999density} optimization of excited states can also be performed as described previously for the ground state~\cite{pham2024direct}. 

The optimization of multiple excited states is performed state-by-state. %This effectively prevents variational collapse and ensures distinct state separation throughout the optimization process. 
The loss functional minimized in the VM TIDFT procedure for electronic state $I$ is a sum of the \emph{intra}state term $\mathcal{E}^{I}$ and the \emph{inter}state penalty term $\Omega_{P}$
\begin{equation} \label{eq:loss}
\begin{split}
\Omega &= \mathcal{E}^{I} + \Omega_{P}
\end{split}
\end{equation}
The intrastate term includes $E^{I}$ the energy of state $I$ and the intrastate penalty that ensures linear independence of MOs of state $I$
\begin{equation} \label{eq:intra}
\begin{split}
\mathcal{E}^{I} &= E^{I} - c^{I}_{p} \sum_{\tau=\alpha,\beta} \ln \det \left( \sigma_{II\tau} \sigma_{II\tau d}^{-1} \right)
\end{split}
\end{equation}
where $\sigma_{II\tau}$ is the overlap matrix for the occupied orbitals of spin $\tau$, and $\sigma_{II\tau d}$ is its diagonal part. The intrastate term was described in our previous treatment of the simple ground-state orbital optimization method~\cite{pham2024direct}.

The interstate term is designed to keep different electronic states orthogonal to each other
\begin{equation} \label{eq:inter}
\begin{split}
\Omega_{P} &= - C_{P} \sum_{\tau=\alpha,\beta} \ln \det (\Phi_{\tau} \Phi^{-1}_{\tau d}) 
\end{split}
\end{equation}
In the state-by-state optimization, the orthogonality penalty is imposed between the currently optimized state and all previously optimized states. The key element of the penalty is $\Phi_{\tau}$ -- the overlap matrix between electronic states. $\Phi_{\tau d}$ is a diagonal matrix containing the elements of $\Phi_{\tau}$. Element $\left(\Phi_\tau \right)_{IJ}$ of the state overlap matrix can be calculated as the determinant of the overlap of MOs that describe states $I$ and $J$
\begin{equation} \label{eq:Phi}
\begin{split}
\left(\Phi_{\tau} \right)_{IJ} &= \det \left( \sigma_{IJ\tau} \right)
\end{split}
\end{equation}
where the MO overlap matrix of occupied orbitals
\begin{equation} \label{eq:sigma} 
(\sigma_{IJ\tau})_{ij}= \sum_{\mu\nu}^{B} T^{I\tau}_{\mu i} S_{\mu \nu} T^{J\tau}_{\nu j}
\end{equation} 
is written using coefficients $T^{I\tau}_{\mu i}$ that represent occupied MOs $\phi_i^{I\tau} (\mathbf{r})$ in terms of $B$ basis set functions $\chi_\mu (\mathbf{r})$
\begin{equation} \label{eq:T} 
\phi_i^{I\tau} (\mathbf{r}) = \sum_{\mu}^{B} \chi_\mu (\mathbf{r}) T^{I\tau}_{\mu i}
\end{equation}
Here, $S_{\mu \nu}$ is the overlap between basis set functions $\mu$ and $\nu$
\begin{equation} \label{eq:S} 
S_{\mu \nu} = \int \chi_\mu(\mathbf{r}) \chi_\nu(\mathbf{r}) d\mathbf{r}
\end{equation}

For non-zero orbitals, the positive determinant in the penalty term in Eq.~(\ref{eq:inter}) is equal to zero when there is linear dependency between electronic states. $\Phi_{\tau d}$ is the normalization factor that ensures the determinant in the penalty functional does not exceed 1. The determinant achieves its highest value of 1 only when all electronic states are orthogonal. With $ \det (\Phi_\tau \Phi_{\tau d}^{-1})$ always kept in the $(0,1]$ interval, the logarithm function $\Omega_P$ lies in the $(+\infty,0]$ interval, making the linearly dependent states and states with orbitals of zero norm inaccessible in the variational procedure, for any $C_P > 0$. A similar penalty term has proven effective to prevent orbital collapse during the SCF optimization of ground-state MOs~\cite{pham2024direct} and during the localization of nonorthogonal occupied and virtual orbitals~\cite{luo2020direct, luo2021variable}.

The key benefit of introducing the interstate penalty term is that it allows to use molecular orbital coefficients -- arguably the most natural descriptors of the electronic degrees of freedom -- as independent variables in the optimization procedure. As shown below, the expressions for the analytical gradient of the loss functional wrt these variables are very simple and can be written in a closed form, making it easy to use a multitude of well-developed unconstrained optimization algorithms, including conjugate gradient algorithms, quasi-Newton algorithms, or trust-region algorithms~\cite{nocedal2006numerical}. 

The gradient of the intrastate term with respect to orbital coefficients is described in our previous work:
\begin{equation} \label{eq:E_grad} 
\begin{split}
G_{\mu i}^{I\tau \mathcal{E}} &\equiv \dfrac{\partial \mathcal{E}^{I}}{\partial T^{I\tau}_{\mu i}}  \\
&= 2 [(I-SP^{I\tau})F^{I\tau}T^{I\tau}(\sigma_{II\tau}^{-1})]_{\mu i} - \\
&-2 c^I_{p} [ST^{I\tau} (\sigma_{II\tau}^{-1} - \sigma_{II\tau d}^{-1})]_{\mu i}
\end{split}
\end{equation}
% - \dfrac{\partial \Omega_{p}}{\partial T^{I\tau}_{\mu i}} 
%
where $F^{I\tau}$ and $P^{I\tau}$ are the Kohn-Sham Hamiltonian and density matrices, respectively, for state $I$ and spin $\tau$. The density matrix is evaluated taking into account nonorthogonality of occupied orbitals:
\begin{equation} \label{eq:define-BS} 
P^{I\tau}_{\mu\nu} = \sum_{ij}^{N_\tau} T^{I\tau}_{\mu i} (\sigma_{II\tau}^{-1})_{ij} T^{I\tau}_{\nu j}
\end{equation}

The first derivatives of the interstate penalty term $\Omega_{P}$ can also be evaluated analytically and computed readily
\begin{equation} \label{eq:P_grad} 
\begin{split}
G_{\mu i}^{I\tau P} & \equiv \dfrac{\partial \Omega_{P}}{\partial T^{I\tau}_{\mu i}} = \\
 & - 2C_{P} \left[ \sum_{J} (\Phi_\tau )^{-1}_{IJ} \left[ ST^{J\tau} \text{adj}(\sigma_{IJ\tau}) \right]_{\mu i} - (ST^{I\tau} \sigma_{II\tau}^{-1})_{\mu i} \right]
\end{split}
\end{equation}
where $\text{adj}(\sigma_{IJ\tau})$ is the adjugate of the singular or nearly-singular (i.e. ill-conditioned) overlap matrix between orbitals of states $I$ and $J$. This adjugate matrix emerges from differentiating the determinant of the overlap matrix (see Supporting Information). 

\label{subsec:algorithm}

\textbf{Optimization algorithm.} In this work, VM TIDFT methods is implemented using the basic preconditioned conjugate gradient (PCG) algorithm for nonlinear problems~\cite{hager2006survey}. To minimize a function, PCG needs only its first derivative (i.e. gradient). Computation of the exact second derivative is not required, but approximate second derivative has been found useful for the construction of accurate preconditioners, which increase the rate of convergence of the PCG algorithm and making it practical.

The analytical gradient of the intrastate term in Eq.~(\ref{eq:E_grad}) can be computed efficiently as done in our previous work. However, the analytical gradient of the interstate term requires the evaluation of the lesser known adjugate matrix of $\sigma_{IJ\tau}$ -- the overlap between MOs of two electronic states. Fortunately, the adjugate of a matrix is well-conditioned even if the matrix itself is ill-conditioned~\cite{stewart1998on}. This property is very important in our case because nearly orthogonal electronic states, which are often encountered during the optimization, produce ill-conditioned $\sigma_{IJ\tau}$.

Our algorithm for computing the adjugate is designed to avoid ill-conditioned intermediates~\cite{stewart1998on}. It relies on computing the singular value decomposition of the overlap
\begin{equation} \label{eq:adj-svd} 
\begin{split}
\text{adj} (\sigma) & = \text{adj}(U d V^\dagger) = \text{adj}(V^\dagger) \text{adj}(d) \text{adj}(U) = \\
&= \det (V) \det(U) V \text{adj}(d) U^\dagger
\end{split}
\end{equation}
and then evaluating the adjugate of the diagonal matrix $d$ using the relation between the co-factor of a matrix and its adjugate
\begin{equation} \label{eq:adj-cofactor} 
\begin{split}
[ \text{adj} (d) ]_{ii} & = [ C^\dagger ]_{ii} = [ \prod_{j\neq i} d_{jj} ]_{ii}
\end{split}
\end{equation}
where $C$ is the diagonal cofactor matrix of $d$. Note that the inversion of singular values is not required in this algorithm, making the algorithm stable for the often encountered ill conditioned matrices.

The preconditioner is constructed by simplifying the second derivative of the loss functional: the interstate term is neglected and the remaining intrastate term is approximated as described in the VM SCF optimization of a single electronic state~\cite{pham2024direct}. 
\begin{equation} \label{eq:preconditioner}
\begin{split}
H^{I\tau}_{\mu \nu}
	&= 4 \left[ (1-\kappa) \left( A^{I\tau\dagger} F^{I\tau} A^{I\tau} \right)_{\mu\nu} \right. + \\
	&+ \left. \kappa \vert \varepsilon_{\text{HOMO}} \vert \left( D^{I\tau\dagger} S D^{I\tau} \right)_{\mu\nu} \right]
\end{split}
\end{equation}
where $\kappa$ is a dimensionless regularization parameter that allows a user to increase the weight of the overlap in case of poor convergence~\cite{pham2024direct}. Matrices $A^{I\tau}$ and $D^{I\tau}$ can be set to the projector onto the unoccupied subspace $Q^{I\tau}\equiv (I - P^{I\tau}S)$, identity matrix $I$ or zero.
This preconditioner is the same for all MOs of an electronic state and can be inverted independently for each state. The numerical tests described in the next section show that this preconditioner allows to optimize excited state MOs efficiently.

The computational cost of the PCG algorithm presented here is determined by both the number of the electronic states $S$ and by the size of the system, that is, by the number electrons $N$, which is prportional to the number of basis set functions $B$. The cost of treating all $S$ intrastate terms is dominated by the cost of inverting $S$ preconditioners, which has $\mathcal{O}(SB^3)$ computational complexity~\cite{pham2024direct, vandevondele2003efficient}. 
It should be noted that the preconditioners are inverted only in the beginning of the PCG procedure, making the direct optimization more efficient compared to the diagonalization-based methods, in which such costly operations are required on each iteration. It should also be noted that the intrastate cost of handling nonorthogonal MOs is lower -- $\mathcal{O}(SB^2 N)$ to compute orbital overlaps $\sigma_{II}$ and $\mathcal{O}(SN^3)$ to invert them -- because the number of electrons $N$ is typically an order of magnitude lower than $B$. The determinants of the intrastate overlap matrices can also be computed relatively fast via their LU decomposition at $\mathcal{O}(SN^3)$ cost. 

Handling the interstate term and its derivative requires $\mathcal{O}(S^2 B^2 N)$ time to build the  $S(S-1)/2$ overlap matrices for all pairs of electronic states, $\mathcal{O}(S^2 N^3)$ time to evaluate their adjugates and determinants (the latter being the elements of matrix $\Phi$), and $\mathcal{O}(S^3)$ time to compute the determinant of $\Phi$. Since the number of electronic states of interest is small compared to $B^2 N \gg S$ for most systems, building the overlap matrices will dominate the interstate computations.
%Number of single excitations is S_1 ~ N(B-N),  S_2 ~ S_1 * (N-1)*(B-N-1) = N(B-N)(N-1)(B)=N^2B^2 

To summarize, the intrastate and interstate terms are dominated by the $\mathcal{O}(S B^3)$ and $\mathcal{O}(S^2 B^2 N)$, respectively. Therefore, the interstate term will represent the computational bottleneck for a large number of excited state, but does not consume significant computational time for most practical problems with $S<100$.

Initial excited state orbitals are generated using fully optimized canonical ground-state orbitals. The guess is obtained by systematically generating all possible excitations involving a fixed number  electrons, typically two, and then retaining only the lowest (unoptimized) energy excitations for further optimization.

\label{subsec:penalty}

\textbf{Penalty strength adjustment.} The interstate penalty term plays a critical role in preventing the collapse of high-energy states onto lower-energy states. Increasing the strength of this term by increasing $C_P$ pushes the electronic states to become closer to being orthogonal.
If electronic states are not completely orthogonal, the optimized excited states can contain noticeable undesirable admixtures of the lower-lying states of the same symmetry.
To produce orthogonal electronic states without strictly enforcing the orthogonality constraint, the interstate penalty strength is gradually increased using a fixed multiplicative update factor $f_{upd} > 1$  
\begin{equation} \label{eq:CP}
\begin{split}
C_{P}^{(n+1)} = C_{P}^{(n)} \times f_{upd}
\end{split}
\end{equation} 
until the electronic states become orthogonal, that is, until the deviation from the orthogonality (DFO) measured by the max norm of the off-diagonal elements of the interstate overlap matrix drops below a user-specified threshold $\mathcal{E}_0$:
\begin{equation} \label{eq:residual}
\begin{split}
\text{DFO} \equiv \lvert\lvert \Phi_{\tau d}^{-1/2}\Phi_{\tau}\Phi_{\tau d}^{-1/2} - I \rvert\rvert_{\text{max}} < \mathcal{E}_0
\end{split}
\end{equation} 
$C_P$ is adjusted in the outer loop, each iteration of which runs the PCG optimizer with a fixed penalty strength. This procedure guarantees that the optimization does not only produces accurate fully optimized MOs for each state, but also enforces orthogonality between the electronic states. 

\label{subsec:details}

\textbf{Computational details.} The VM TIDFT procedure was implemented in the electronic structure module of the CP2K software package~\cite{thomas2020cp2k}. In the current implementation, matrix operations were performed using the DBCSR~\cite{borstnik2014sparse} library for sparse matrices and ScaLAPACK~\cite{choi1996scalapack} library for dense matrices. The default values for the input parameters that control convergence of the PCG VM optimization are shown in Table~\ref{table:input} and used unless specified otherwise.

\begin{table*}[htb] %[ht!]
	%\centering
	\begin{tabular}{l c c c} 
		\hline
		Parameter & Notation & Default value & Described in \\
		\hline
		%Initial intrastate penalty strength & $C^{I}_p$ & $10^{-5}~\text{Ha}$ & Ref.~\cite{pham2024direct} \\
		Initial interstate penalty strength & $C_P^{(0)}$ & $10~\text{Ha}$ & Eq.~(\ref{eq:CP}) \\
		%Number of steps in the $C_p$ loop & $N$&  3 &  Ref.~\cite{pham2024direct} \\
		SCF convergence threshold & $\lvert\lvert \frac{\partial \Omega}{\partial T}\rvert\rvert_{\text{RMS}}$ & $10^{-6}~\text{Ha}$ &  \\
		%Initial loose SCF threshold& $\epsilon_{0}$& $10^{-2}~\text{Ha}$& Ref.~\cite{pham2024direct} \\
		Final allowed DFO & $\mathcal{E}_0$ & $10^{-3}$ & Eq.~(\ref{eq:residual}) \\
		Interstate update factor & $f_{upd}$ & 2 & Eq.~(\ref{eq:CP}) \\
		%Maximum range of guess excitation & $E_{guess-max}$ & 1000 e.V & Fig.~\ref{RZZK} \\
		%Minimum range of guess excitation & $E_{guess-min}$ & 0.0 e.V & Fig.~\ref{RZZK} \\
		Hamiltonian term in the preconditioner& $A$& 0 & Eq.~(\ref{eq:preconditioner}) and Ref.~\cite{pham2024direct} \\
		Overlap term in the preconditioner& $D$& $I$& Eq.~(\ref{eq:preconditioner}) and Ref.~\cite{pham2024direct} \\
		\hline
	\end{tabular}
	\caption{Input parameters that control the PCG optimization of MO coefficients.}
	\label{table:input}
\end{table*}

%\begin{figure*}[htb] %tb
%	\includegraphics[width=0.45\textwidth]{figures/test_systems.pdf}
%	\caption{Organic test systems.}
%\label{fig:tests}
%\end{figure*}

The efficiency and accuracy of the VM TIDFT method was tested on systems including gas-phase atoms (He, Be, B, Li) and small molecules (azulene, benzene, butadiene, ethene, formaldehyde, hydrogen chloride, hydrogen fluoride, nitrobenzene, N-phenylpyrrole, pyrrole, pyrazine, and naphthalene). We also calculated excitations in molecular complexes C$_2$H$_4$\ldots C$_2$F$_4$, NH$_3$-BF$_3$, NH$_3$\ldots F$_2$, and $\pi$-stacked dimer of pyrrole and pyrazine molecules. The organic molecules, particularly azulene, benzene, butadiene, nitrobenzene, and N-phenylpyrrole, are of special interest due to their distinct excited-state behavior driven by CT and bonding-antibonding interactions. These conjugated systems, with near-degenerate core and delocalized $\pi$-orbitals, are prone to root-flipping issues~\cite{marie2023excited, tran2019tracking} during excited-state calculations, complicating the identification of states and making them excellent benchmarks. Moreover, the availability of extensive experimental and computational data for these systems enables meaningful comparisons, ensuring the reliability and accuracy of the tested method.

Unrestricted Kohn-Sham formalism (UKS) was used to describe spin-orbitals in this study unless indicated otherwise. 
Test calculations were carried out using the %dispersion corrected~\cite{grimme2010dispersion}
generalized gradient exchange-correlation functional of Perdew, Burke and Ernzerhof (PBE)~\cite{zhang1998comment, perdew2008restoring, perdew1997generalized}. In the dual Gaussian and plane-wave scheme implemented in CP2K~\cite{hutter2014cp2k}, double-$\zeta$ valence basis set with one set of polarization functions (DZVP)~\cite{vandevondele2007gaussian} were used to represent atomic orbitals. A plane-wave cutoff of 300~Ry was established to adequately describe the electron density, whereas a higher cutoff of 1500~Ry was used to produce smooth dissociation curves for the hydrogen $H_2$ and the HeH$^+$ molecules. Separable norm-conserving Goedecker-Teter-Hutter pseudopotentials were used to describe the interactions between the valence electrons and ionic cores~\cite{goedecker1996separable, krack2005pseudopotentials}. All calculations were performed using a cubic simulation cell with dimensions of $15\times 15\times 15$ for $H_2$ and $30\times 30\times 30$ for the rest of the systems. Electrostatic interactions between the periodic cell was eliminated using wavelet approach~\cite{genovese2006efficient, genovese2007efficient}.

\section{Data Availability}

Optimized atomic coordinates for all structures and lattice vectors are deposited in a Figshare database.%~\cite{pham2024directes}.

\section{Results and Discussion}

\label{subsec:penalty-rnd}

\begin{figure}[htbp]
	\includegraphics[width=0.45\textwidth]{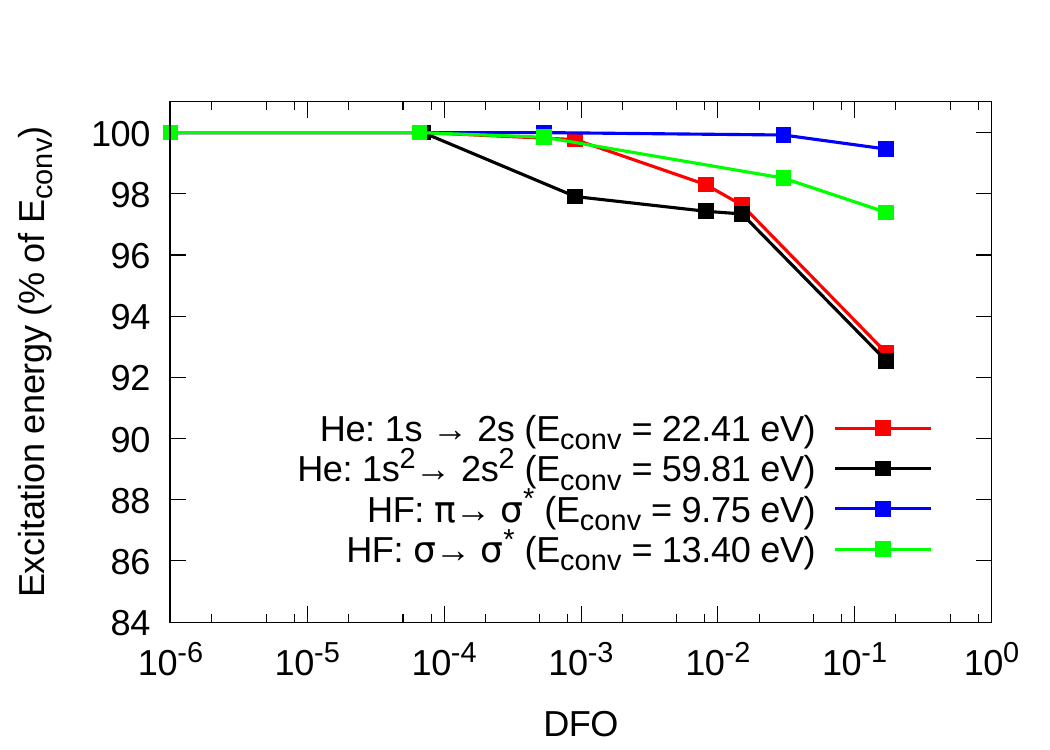}
	\caption{Excitation energy for the first two excitations of He atom (PBE/TZVP) and HF molecule (PBE/DZVP) computed with VM TIDFT optimization procedure for the DFO values measured after convergence is reached. The recovered excitation energy is a fraction of the converged excitation energy $E_{\text{conv}}$ obtained with very tight requirement for the orthogonality criterion ($\mathcal{E}_0=10^{-5}$).}
\label{fig:HF_dev}
\end{figure}

\textbf{Penalty strength effect.} Fig.~\ref{fig:HF_dev} shows the impact of the residual nonorthogonality between electronic states on the excitation energies. With higher penalty strength $C_P$, the electronic states are pushed to become orthogonal decreasing DFO and recovering the true excitation energies. When the allowed DFO is too large, only a fraction of the true excitation energies is recovered, signalling the admixture of lower-lying states due to insufficient orthogonality enforcement. This data highlights the importance of the interstate penalty term and suggests that $\mathcal{E}_0 = 10^{-4}$ is an acceptable DFO that recovers a large fraction of excitation energies. For HF, the bonding interaction arises from the overlap between the hydrogen 1s orbital and the fluorine 2p$_z$ orbital, while the 1$\pi$ orbitals are non-bonding as they have wrong symmetry to interact with the H 1s orbital and are localized primarily on the fluorine atom.

\label{subsec:accuracy}

%RZK after thesis submission: Discuss accuracy of the higher-energy states.
%RZK after thesis submission: Challenging excitations for which TDDFT fails completely must also be included.
%RZK after thesis submission: Better comparison to OCDFT must be made: results should be spin-purified and def2-TZVP basis should be used.
\textbf{Accuracy.} Table~\ref{tab:example} compares experimentally measured first singlet excitation energies with those computed using VM TIDFT, TDDFT in Tamm-Dancoff approximation, OCDFT, and wavefunction-based correlation methods (MP2 or CCSD). 

%RZZK: Are OCDFT excited state wavefunctions spin-purified? Is it discussed in OCDFT article?
%CINDY: Yes, they proposed 2 procedures for spin-adaptation. They reported the excitations using these 2 procedures in SI and I took the data with spin-raising correction procedure in table S1. 
Despite different optimization paradigms, VM TIDFT utilizes the same orthogonality constraint as OCDFT. Therefore, the fully optimized states in VM TIDFT and OCDFT are expected to be the same if computed using the same model chemistry (i.e. the same exchange-correlation functional and basis set). Table~\ref{tab:example} shows that VM TIDFT excitation energies are indeed close to those computed with OCDFT but not the same. This is due to the OCDFT results being partially spin-purified in the post-SCF procedure~\cite{evangelista2013orthogonality}, whereas spin-purification was not used in VM TIDFT optimization. The results are also different due to the different basis sets employed to represent occupied states: def2-TZVP in OCDFT is a larger more flexible basis set than DZVP in VM TIDFT. 

For the six systems with complete experimental and computational data (first six rows in Table~\ref{tab:example}), VM TIDFT results are in reasonable agreement with experimental data ($\text{MAD} = 1.02$~eV). For comparison, the partially spin-purified OCDFT results are in slightly better agreement with experimental data than VM TIDFT ($\text{MAD} = 0.89$~eV), whereas TDDFT agreement is significantly worse ($\text{MAD} = 1.71$~eV).

\begin{figure}[htbp]
\centering
	\includegraphics[width=0.45\textwidth]{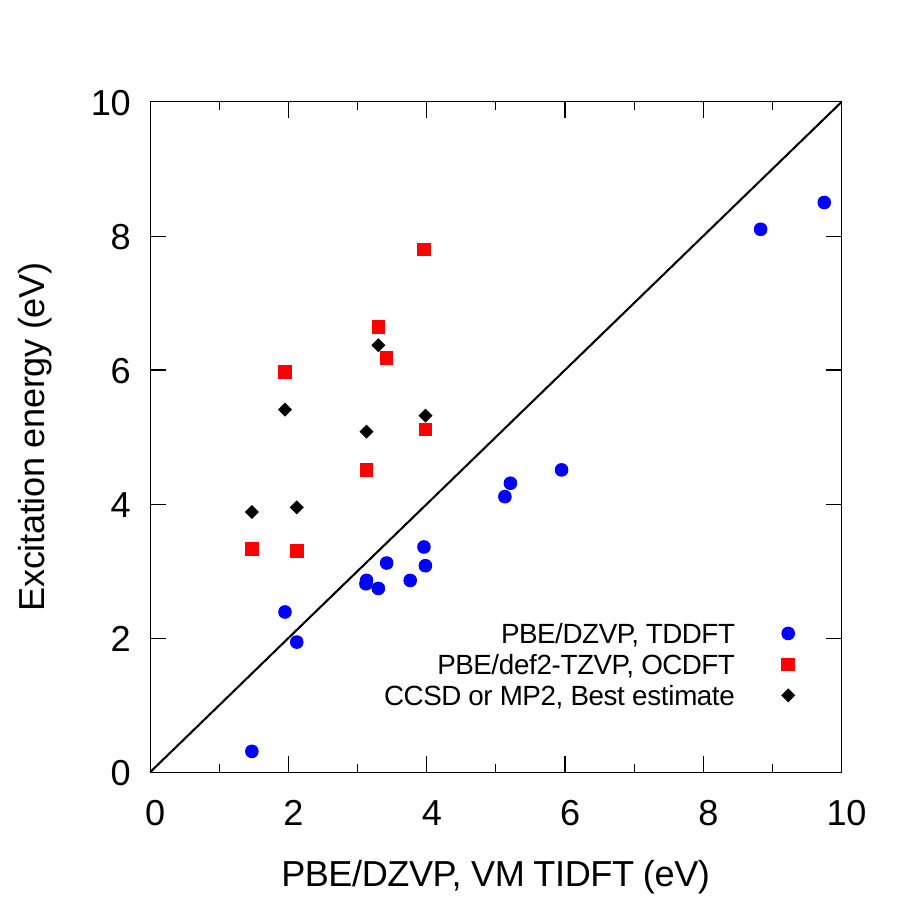}
	\caption{Comparison of VM TIDFT excitation energies with those calculated with TDDFT, OCDFT, and wavefunction correlation methods. TDDFT and VM TIDFT results were obtained in this work using CP2K, OCDFT  energies were computed using Psi4 in Ref.~\onlinecite{evangelista2013orthogonality}, and MP2/6-31G* and CCSD/aug-cc-pVTZ results were computed using Gaussian in Ref.~\cite{schreiber2008benchmarks}.}
\label{fig:compare1}
\end{figure}

When all test systems with available experimental excitation energies, are compared (not just the first six of them), the accuracy of VM TIDFT ($\text{MAD} = 0.88$~eV) remains noticeably better than that of TDDFT ($\text{MAD} = 1.36$~eV). Comparison between VM TIDFT and TDDFT shows that TDDFT tends to slightly underestimate excitation energies for the test systems (Fig.~\ref{fig:compare1}). On the other hand, VM TIDFT energies are underestimated compared to OCDFT, MP2, and CCSD energies (Fig.~\ref{fig:compare1}). OCDFT data suggests that using spin-purified VM TIDFT wavefunctions, described with larger basis set can substantially improve the excitation energies. Spin-purified VM TIDFT calculations will be performed using the ROKS optimizer in the future.

It is worth noting the magnitude of the residual norm of the energy gradient for the fully optimized excited states (Table~\ref{tab:example}). For most excited states, this norm is below the convergence threshold of $10^{-6}$~Ha imposed on the loss (not the energy) functional, indicating that these excited states are true minima of the DFT functional. Once optimized, these states will not collapse to the lower lying ground state even without the orthogonality constraint. There are, however, several excited systems with the energy gradient higher than the $10^{-6}$~Ha threshold. These are He atom, N-phenylpyrrole, NH$_3\ldots$BF$_3$, and pyrrole-pyrazine dimer. In these systems, the first excited states are not electronic minima and, in these optimized states, the non-zero energy gradient is fully compensated by the opposite non-zero gradient of the interstate penalty functional, producing the zero-gradient true minima of the loss functional. 
It is also important to note that such stationary points still correspond to physically meaningful excited states~\cite{perdew1985extrema}. For instance, VM TIDFT results for the He atom match closely TDDFT results and experimental data for low-lying excitations while not being true electronic minima (Table~\ref{tab:He-states}). This confirms the validity of using orthogonality constraints in orbital-optimized excited-state DFT.

\begin{table*}[htb!]
\centering
\begin{tabular}{ccccc}
\hline
\hline
 &  &  & RMS norm of the &   \\
Excitation           &  TDDFT     &  VM TIDFT     & energy gradient &  Experiment  \\
           & (eV)  & (eV)  & (Ha)            & (eV) \\
 \hline
 $1s^2 \rightarrow 1s2s$ & 17.48  & 20.27    & 7.9$\times 10^{-2}$	& 20.62~\cite{xie2010inelastic}  \\
 $1s^2 \rightarrow 1s2p$ & 21.07  & 22.17    & 9.1$\times 10^{-7}$	& 21.22~\cite{xie2010inelastic} \\
 $1s^2 \rightarrow 1s3s$ & 30.87  & 29.52	   & 1.9$\times 10^{-2}$ & 22.93~\cite{xie2010inelastic} \\
% G16, TDDFT-aug-cc-pVQZ
% 17.6100  - 1s2s
% 22.1342 x 3 - 1s2p 
% 33.2301 - 1s3s
% 41.5713 x 5 - 1s 3d
% 50.7528 x 3 - 1s 3p
% 88.7483 x 7 - 1s 4f?
% $1s^2 \rightarrow 1s2s$ & 20.24  & 20.91    & 9.7$\times 10^{-2}$	& 20.62~\cite{xie2010inelastic}  \\
% $1s^2 \rightarrow 1s2p$ & 40.64  & 36.95    & 3.2$\times 10^{-5}$	& 21.22~\cite{xie2010inelastic} \\
% $1s^2 \rightarrow 1s3s$ & 37.47  & 39.10	   & 5.3$\times 10^{-2}$ & 22.93~\cite{xie2010inelastic} \\
% $1s^2 \rightarrow 1s3p$ & -  & -	   & - & 23.09~\cite{xie2010inelastic} \\
% $1s^2 \rightarrow 1s4s$ & -  & -	   & - & 23.68~\cite{xie2010inelastic} \\
% $1s^2 \rightarrow 1s4p$ & -  & -	   & - & 23.75~\cite{xie2010inelastic} \\
 $1s^2 \rightarrow 2s^2$ &  -     & 58.09    & 4.3$\times 10^{-1}$ & 57.84~\cite{sekikawa2009two}  \\  
 $1s^2 \rightarrow 2s2p$ &  -     & 65.79    & 3.9$\times 10^{-1}$ & 60.15~\cite{aufleger2022line} \\
\hline
\hline
\end{tabular}
\caption{PBE/aug-cc-pV5Z excitation energies calculated using TDDFT and VM TIDFT with $\mathcal{E}_0=10^{-5}$.} 
%RZK: I am afraid I cannot immediately explain the huge discrepancy between experimental and computed excitation energies involving $2p$ and $3s$ orbitals. Most likely the high-energy transitions are not accurate because of the insufficient variational flexibility of the QZV3P basis set. It is not only VM TIDFT failure, but also TDDFT failure. See also parahelium (singlet helium) data:
%https://en.wikipedia.org/wiki/Helium_atom#/media/File:Helium-term-scheme.svg
%RZK: Also note that your aug-cc-pVTZ give much lower 1s->2p excitation energy of 23.84 eV. Although I am not sure EMSL_BASIS_SETS can be used with GTH_POTENTIALS, this basis set will give us much better agrement with experiments. Please re-run He with aug-cc-pVTZ or even aug-cc-pVQZ. Increase Epsilon_0 to  10−5. For comparison, also run G16 TDDFT with the same basis set.
%CINDY: DONE. Results with aug-cc-pV5Z have been updated.
\label{tab:He-states}
\end{table*}

\begin{table*}[htb!]
\centering
\begin{tabular}{ccccccc}
\hline
\hline
	  \multirow{3}{*}{Molecules} & RMS norm of &  \multicolumn{2}{c}{CP2K} & Psi4  & Gaussian & \multirow{3}{*}{Experiment} \\

	& the energy & \multicolumn{2}{c}{PBE/DZVP} & PBE/def2-TZVP & MP2/6-31G*   & \\
	& gradient &							   &			   & & CCSD/aug-cc-pVTZ  & \\
	
\cline{3-4}
\cline{5-5}
\cline{6-6}
 	&  & TDDFT & VM TIDFT & OCDFT~\cite{evangelista2013orthogonality} & Best estimate~\cite{schreiber2008benchmarks}  & \\
 	& (Ha)  &  (eV)	&   (eV)    &  (eV)  & (eV) & (eV)\\
\hline
\hline
pyrazine 			  		& 6.0$\times 10^{-7}$ & 2.86 & 3.13 & 3.30 & 3.95\footref{b} &  3.83~\cite{hong1974lowest} \\

formadehyde 		  		&  8.4$\times 10^{-7}$ & 3.12 & 3.42 & 3.33  & 3.88\footref{b} &  4.28~\cite{moule1975ultraviolet}\\

benzene 			& 9.9$\times 10^{-7}$ & 4.11 & 5.13 & 5.11 & 5.08\footnote{\label{b} MP2/6-31G(d) results~\cite{schreiber2008benchmarks}} & 4.79~\cite{worth2007using}  \\

pyrrole 			  		& 7.8$\times 10^{-7}$ & 4.31 & 5.21 & 5.97 & 6.37\footref{b} & 5.22~\cite{andres1993theoretical} \\

butadiene 			  		& 7.3$\times 10^{-7}$ & 2.86 & 3.76 & 4.51 & 6.18\footref{b} & 5.92~\cite{andres1993towards} \\

ethene				  		& 4.9$\times 10^{-7}$ & 4.51 & 5.95 & 6.64 & 7.80\footref{b} & 8.00~\cite{andres1993towards} \\

\hline

azulene 			& 8.9$\times 10^{-7}$  & 1.94 & 2.12 & - & 3.85\footnote{\label{a} CCSD/aug-cc-pVTZ results~\cite{schreiber2008benchmarks}} & 1.77~\cite{palmer2022excited}\\

B atom 			  		& 5.9$\times 10^{-7}$ & 0.31 & 1.47 & -	 & -	 & 1.86~\cite{lynam1992first} \\

Be atom  	  		& 5.5$\times 10^{-7}$ & 3.08 & 3.98 & - 	& - 	& 2.73~\cite{puchalski2013testing}  \\

Li atom		  		        & 7.2$\times 10^{-7}$ & 2.39 & 1.95 & -    & -   	& 3.37~\cite{radziemski1995fourier} \\

naphthalene 		  		& 9.1$\times 10^{-7}$ & 2.74 & 3.30 & -  	& - 	& 3.97~\cite{behlen1981flurorescence} \\

N-phenylpyrrole 	  		& 4.2$\times 10^{-3}$ & 3.36 & 3.96 & - & 5.32\footref{a} &  4.35~\cite{druzhinin2010intramolecular}\\

nitrobenzene 		  		& 8.9$\times 10^{-7}$ & 2.81 & 3.12 & - & 5.41\footref{a} & 4.38~\cite{nagakura1964electronic}  \\

HCl 				  		& 8.8$\times 10^{-7}$ & 8.10 & 8.83 & -	 & - 	& 9.56~\cite{douglas1979electronic} \\

HF					  		& 6.2$\times 10^{-7}$ & 8.50 & 9.75 & - 	& - 	& 10.33~\cite{lonardo1973elkectronic}\\

He atom 		    		& 1.9$\times 10^{-1}$ & 24.15 & 26.03 & -   & - 	& 20.62~\cite{xie2010inelastic} \\

\hline

NH$_3\ldots$F$_2$ 		  		& 9.1$\times 10^{-7}$ & 1.90 & 2.50 & - & - & - \\

pyrrole$\ldots$pyrazine 					& 1.8$\times 10^{-6}$ &  2.52 & 3.10 & - & - & - \\

C$_2$H$_4\ldots$C$_2$F$_4$ 		& 5.8$\times 10^{-7}$ & 4.39 & 5.46 & -	 & - & - \\

NH$_3$-BF$_3$ 		  		& 1.8$\times 10^{-4}$ & 7.29 & 8.81 & - & - & - \\
		
\hline
\hline

MAE vs Experiment\footnote{\label{c} Computed for the first six systems} &  & 1.71 & 1.02 & 0.89 & 0.52 &  0.00 \\

MAE vs Experiment\footnote{\label{d} Computed for the systems with listed experimental values} &  & 1.36 & 0.88 & - & - & 0.00 \\

%MAE vs VM TIDFT &  & 0.83 & 0.00 & - & - & - \\

\hline
\hline
\end{tabular}
\caption{First singlet excitation energies calculated using TDDFT, VM TIDFT, OCDFT and high-level quantum mechanics calculations and shown alongside experimental data.}
\label{tab:example}
\end{table*}

\label{subsec:2e}
\textbf{Double-electron excitations.} Unlike TDDFT, VM TIDFT can trivially describe double-electron excitations. As an example, the energies calculated for the $1s^2 \rightarrow 2s^2$ and $1s^2 \rightarrow 2s^1 2p^1$ excitations in the helium atom are 57.68~eV and 67.63~eV, respectively, at the PBE/QZV3P level of theory. These values are close to the experimental values for these transitions, which are 57.84~eV~\cite{sekikawa2009two} and 60.15~eV~\cite{aufleger2022line}, respectively. It should be noted that OCDFT with a single electron hole-particle pair cannot describe double-electron excitations~\cite{evangelista2013orthogonality}.

\begin{figure}[htbp]
	\includegraphics[width=0.48\textwidth]{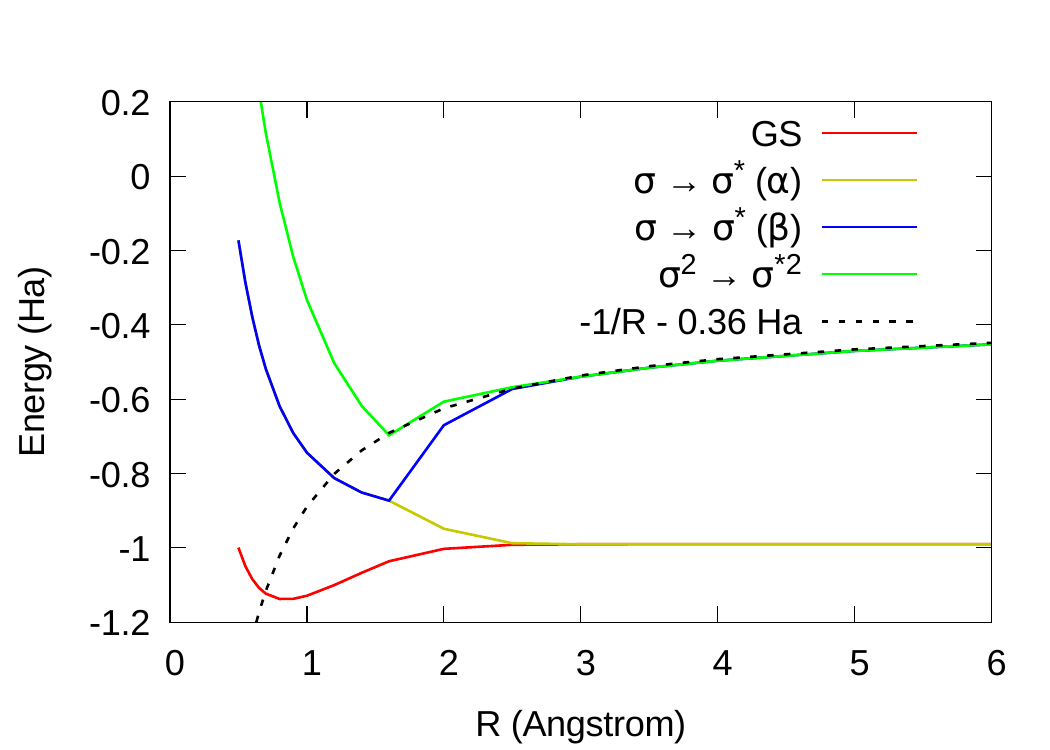}
	\caption{PBE/SZV ground-state and excited-state dissociation curves for H$_{2}$ computed with VM TIDFT. The energy of the Coulomb interactions is down-shifted by 0.36~Ha to account for sum of the ionization potential (IP) and electron affinity (EA) of the isolated hydrogen atoms.}
\label{fig:H2}
\end{figure}

\label{subsec:ct}
\textbf{Charge-transfer excitations.} One of the most important advantages of OODFT methods over TDDFT is their ability to predict the energy of CT excitations accurately. Here, VM TIDFT is used to describe the CT excitation in a minimal-basis H$_{2}$ molecule -- a simple four-level test system previously employed to explain the success of OCDFT and failures of TDDFT~\cite{evangelista2013orthogonality}. Fig.~\ref{fig:H2} shows that VM TIDFT indeed correctly reproduces the long-range behavior of the CT excitation energy, matching the Coulomb potential of the interacting H$^+$ and H$^-$ ions perfectly.  
This behavior can be contrasted with the TDDFT CT energy of the excited states, which shows the attenuation of the Coulomb tail~\cite{evangelista2013orthogonality}, the degree of which is determined by the fraction of the exact exchange in the employed exchange functional.

\begin{figure}[htbp]
	\includegraphics[width=0.48\textwidth]{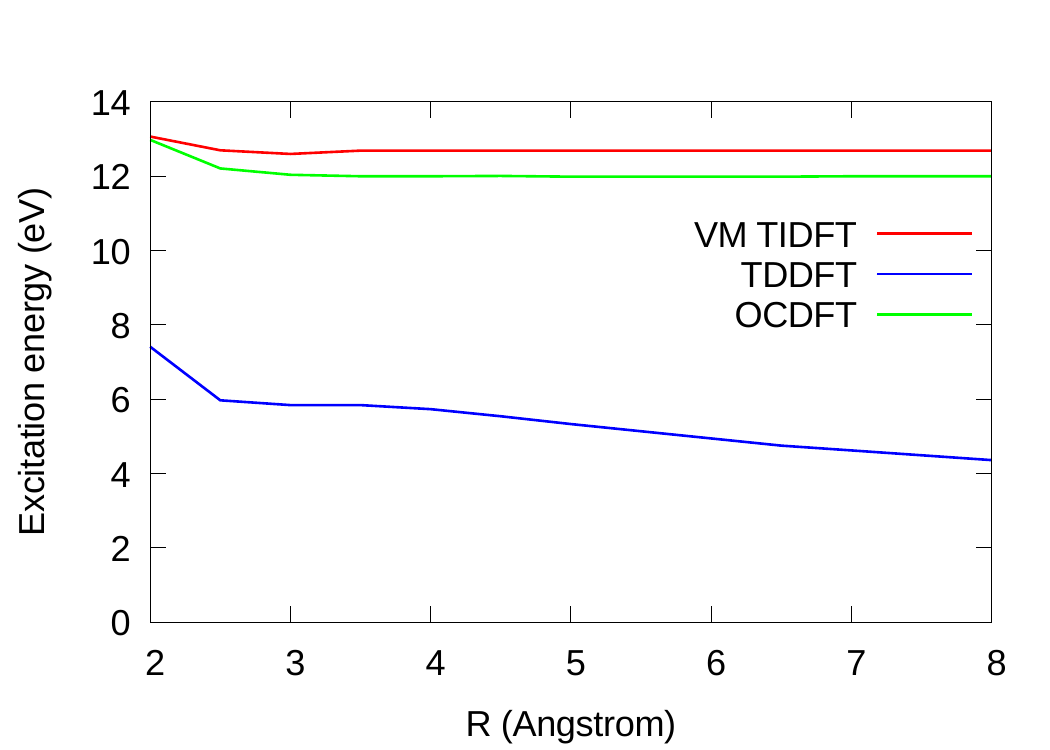}
	\caption{Excitation energy for the charge transfer state of HeH$^{+}$ as a function of the He--H distance, computed using VM TIDFT and TDDFT at the PBE/DZVP level of theory. For comparison, OCDFT PBE/STO-3G results are also shown~\cite{evangelista2013orthogonality}.}
\label{fig:HHe}
\end{figure}

It is also instructive to demonstrate the correct behavior of the CT excitation energies in the [He···H]$^{+}$ ion. For large interatomic distances, the two electrons in the [He···H]$^{+}$ system are localized on the helium atom in the ground state. Upon excitation, one of the electrons transfers from helium to hydrogen, resulting in a He$^{+}\ldots$H configuration. The energy of this excitation is the sum of the ionization potential (IP) of helium and the electron affinity (EA) of hydrogen. Crucially, this energy does not depend on the interatomic distance because the Coulomb interaction between the neutral and positive fragments is zero. VM TIDFT correctly reproduces the flat excitation energy in the large separation limit, in agreement with the spin-purified OCDFT excitation energy (Fig.~\ref{fig:HHe}). In contrast, TDDFT underestimates the excitation energy and fails to capture its asymptotic behavior (Fig.~\ref{fig:HHe}) because of the incomplete cancellation of the Coulomb and exchange interactions between electrons in the PBE and other exchange-correlation functionals.

%\subsection{Rydberg States}
%Low-lying and high-energy Rydberg states discussion.
%, while a quadruple-$\zeta$ valence basis set with three sets of polarization functions (QZV3P) was employed for Rydberg state calculations.

\begin{figure*}[htbp]
  \centering
    \includegraphics[width=0.30\textwidth]{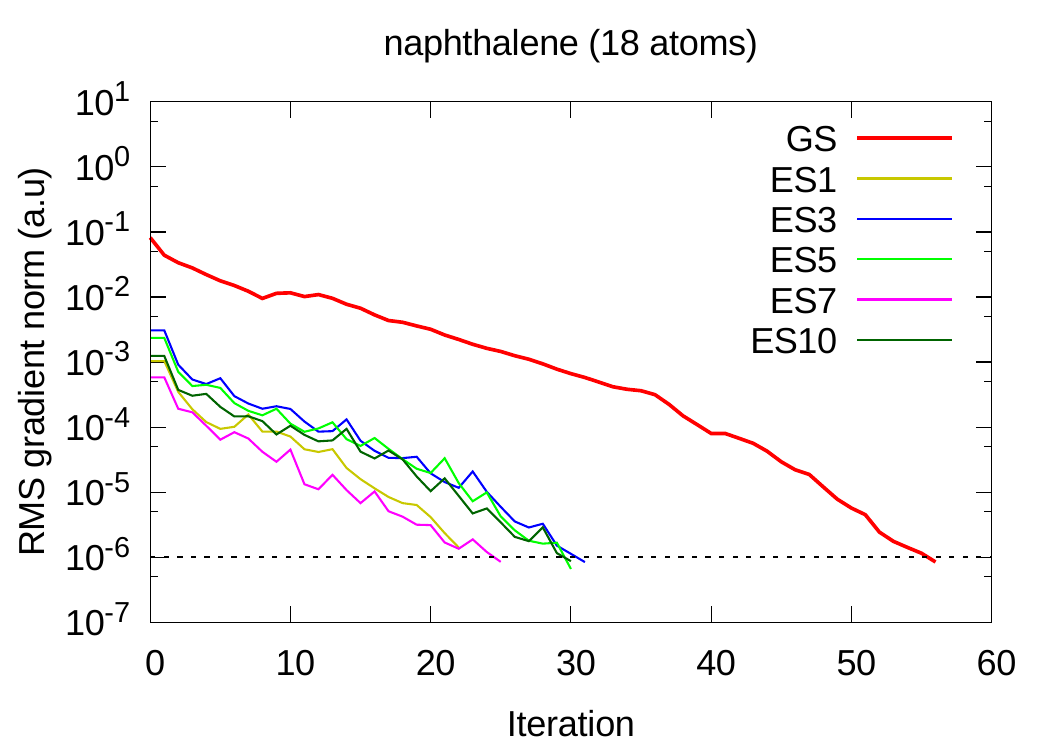}	
    \includegraphics[width=0.30\textwidth]{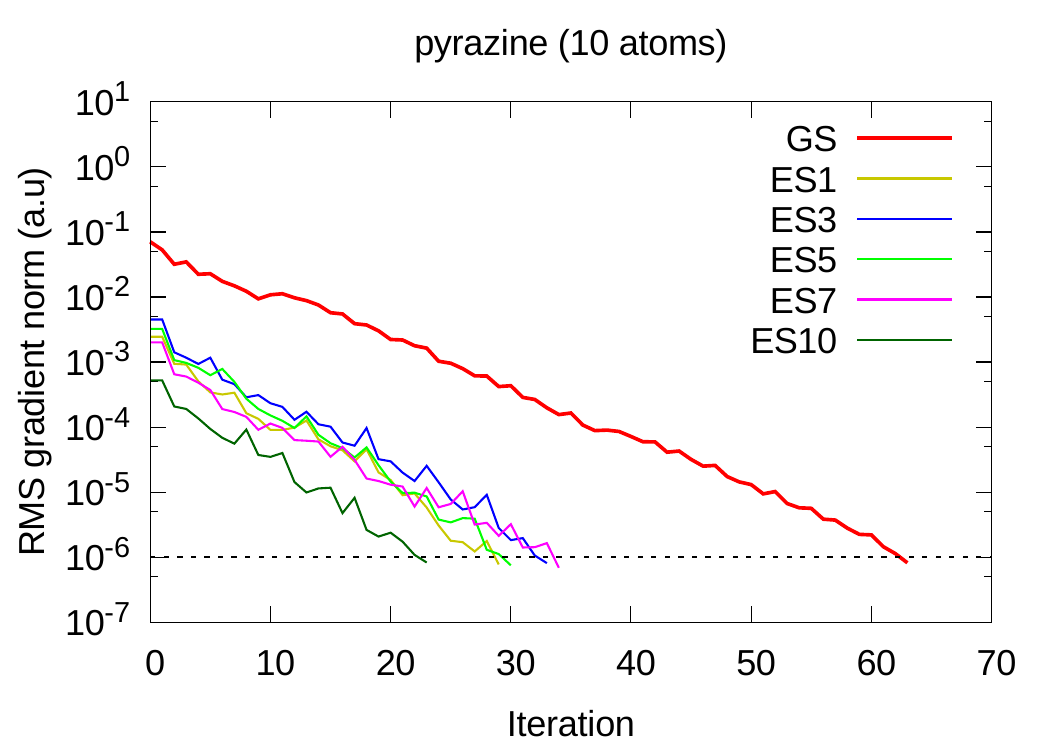}
    \includegraphics[width=0.30\textwidth]{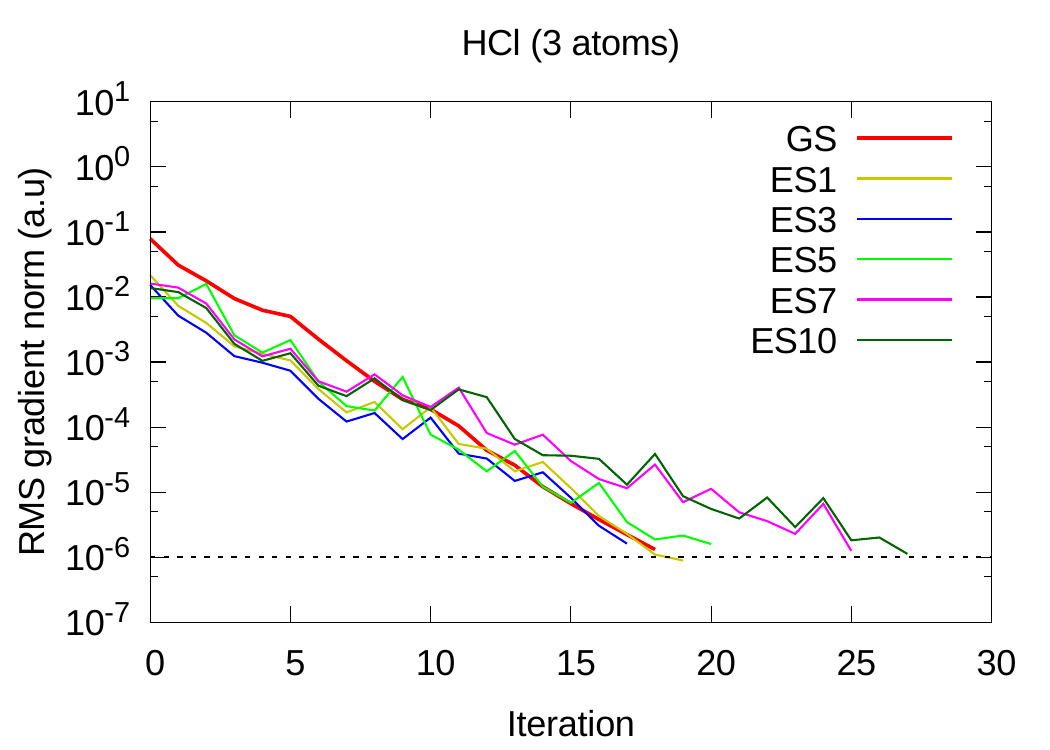}
    \includegraphics[width=0.30\textwidth]{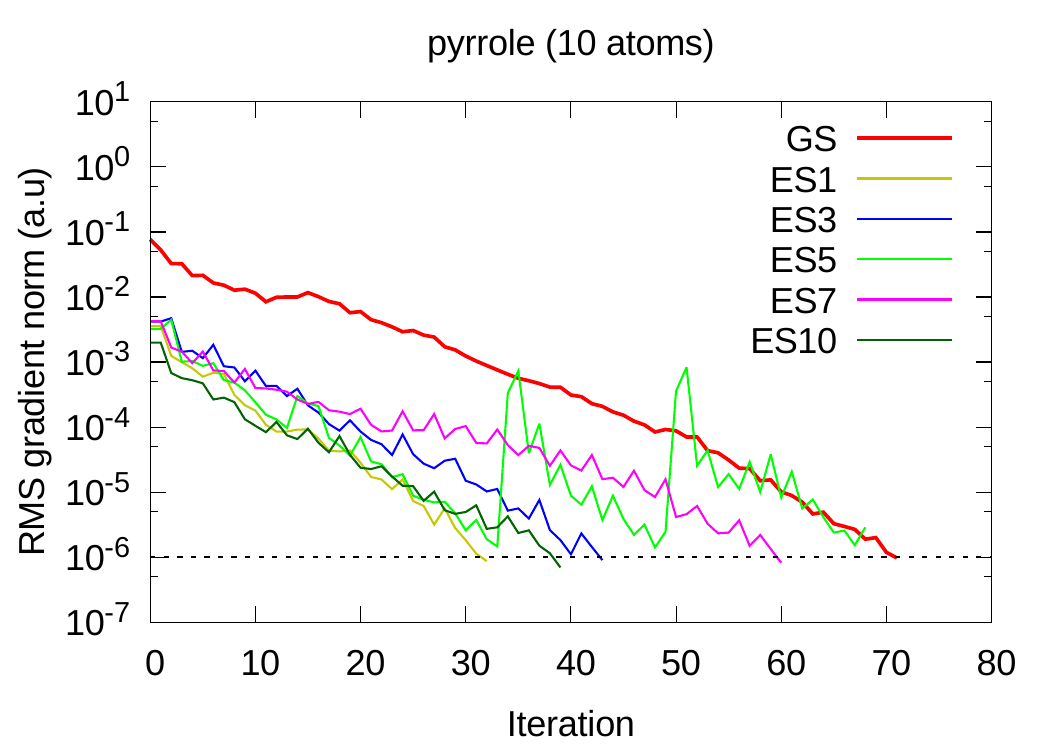}
    \includegraphics[width=0.30\textwidth]{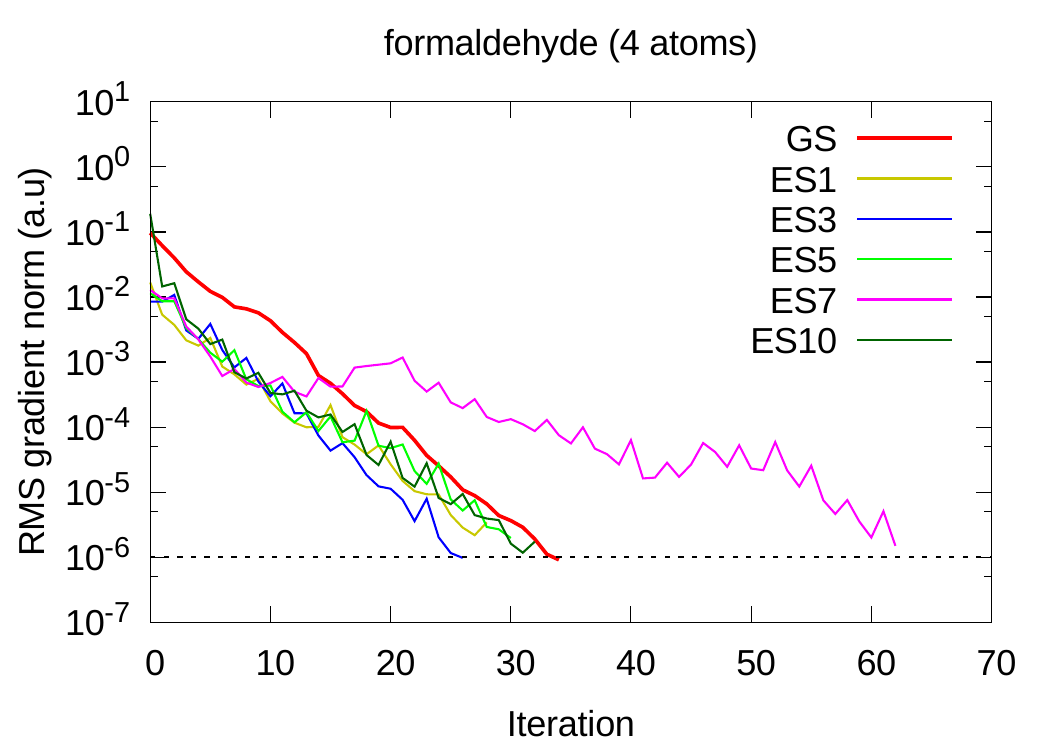}	  
    \includegraphics[width=0.30\textwidth]{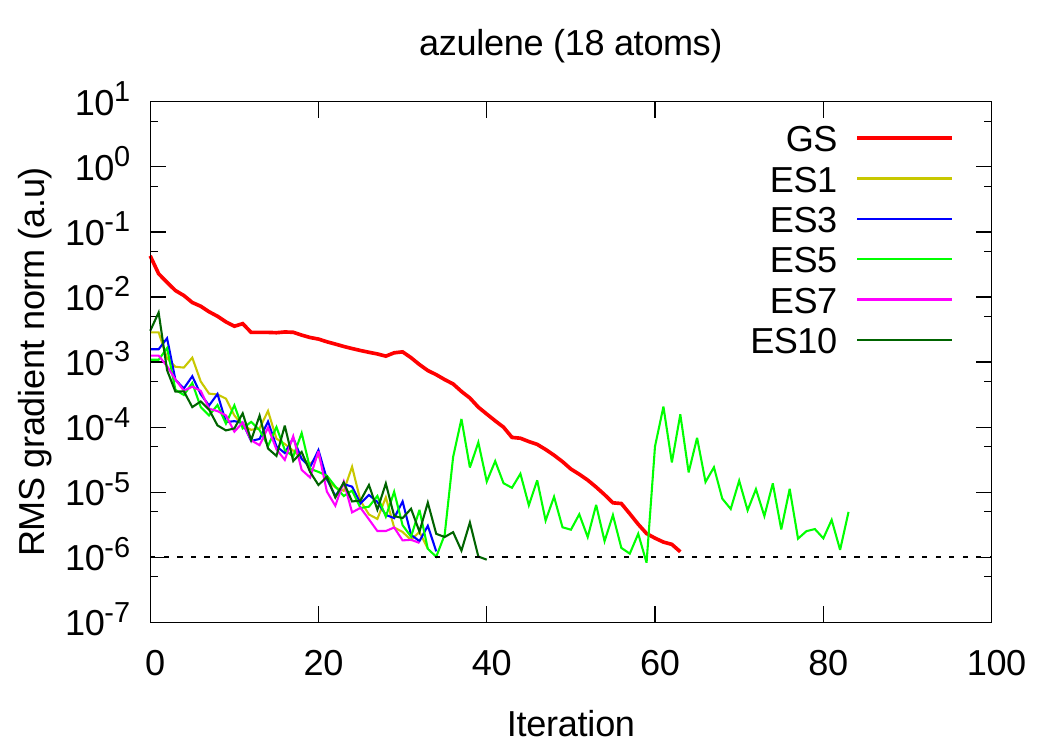}	
    \includegraphics[width=0.30\textwidth]{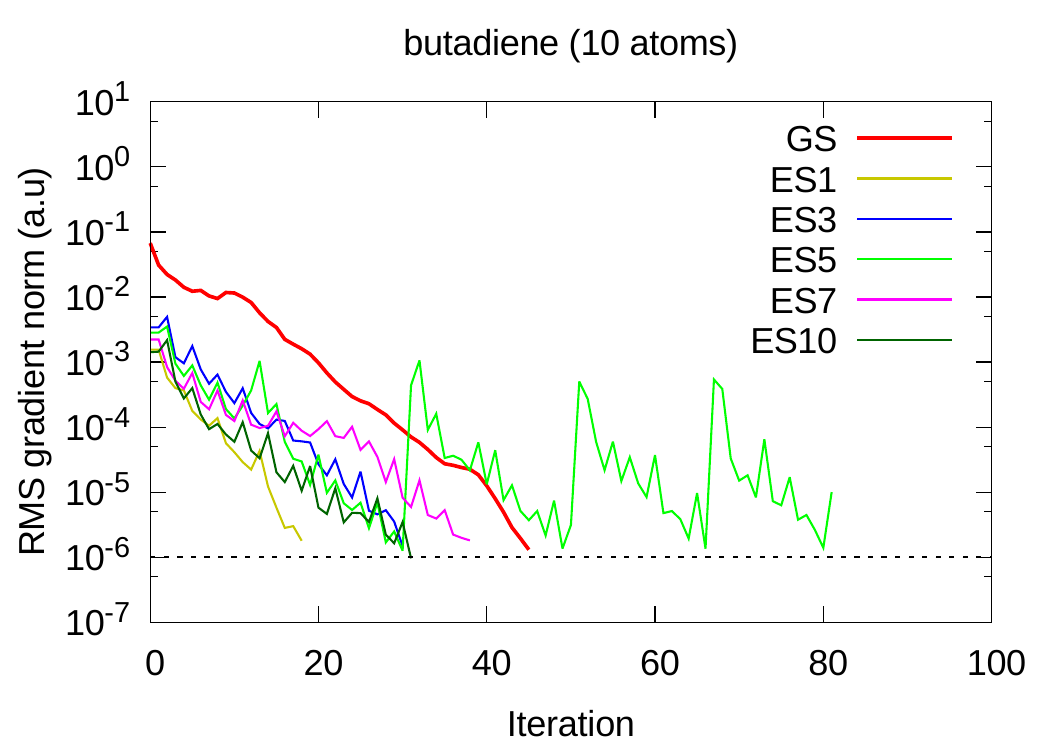} 	
    \includegraphics[width=0.30\textwidth]{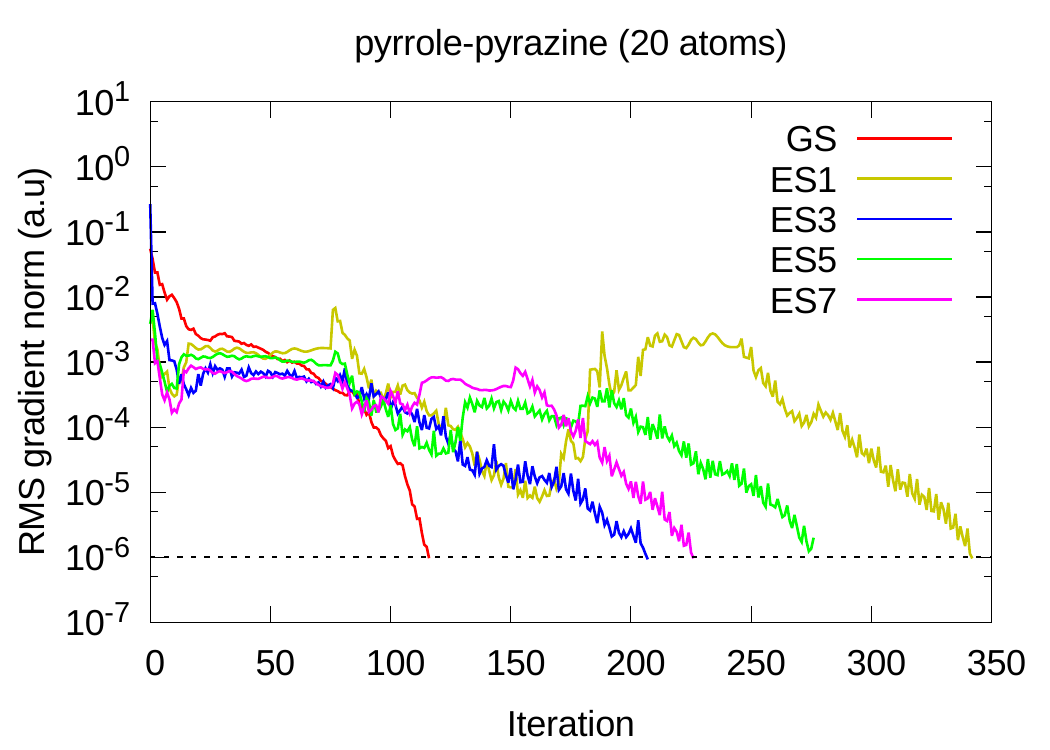}
    \includegraphics[width=0.30\textwidth]{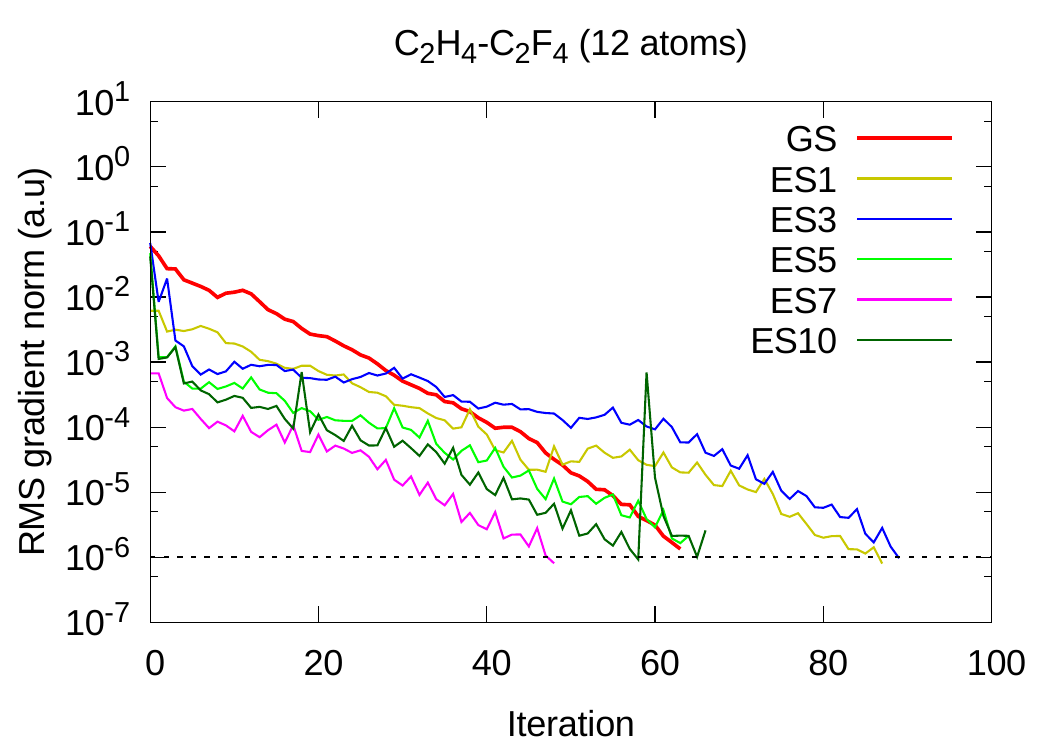}	 
    \includegraphics[width=0.30\textwidth]{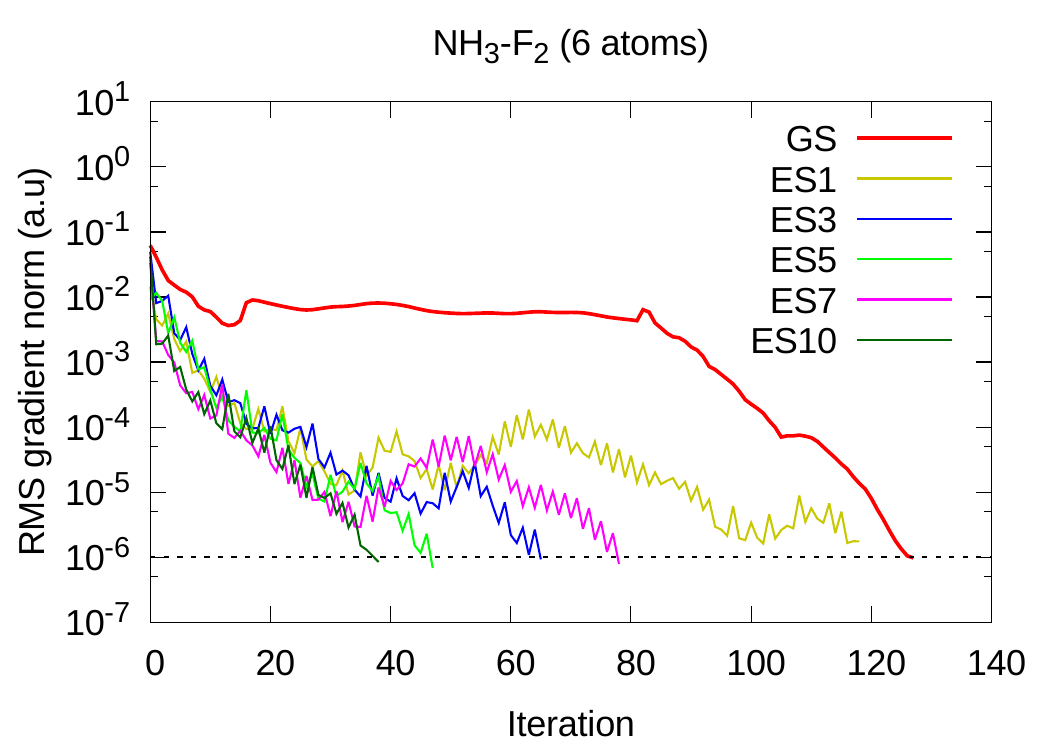}
    \includegraphics[width=0.30\textwidth]{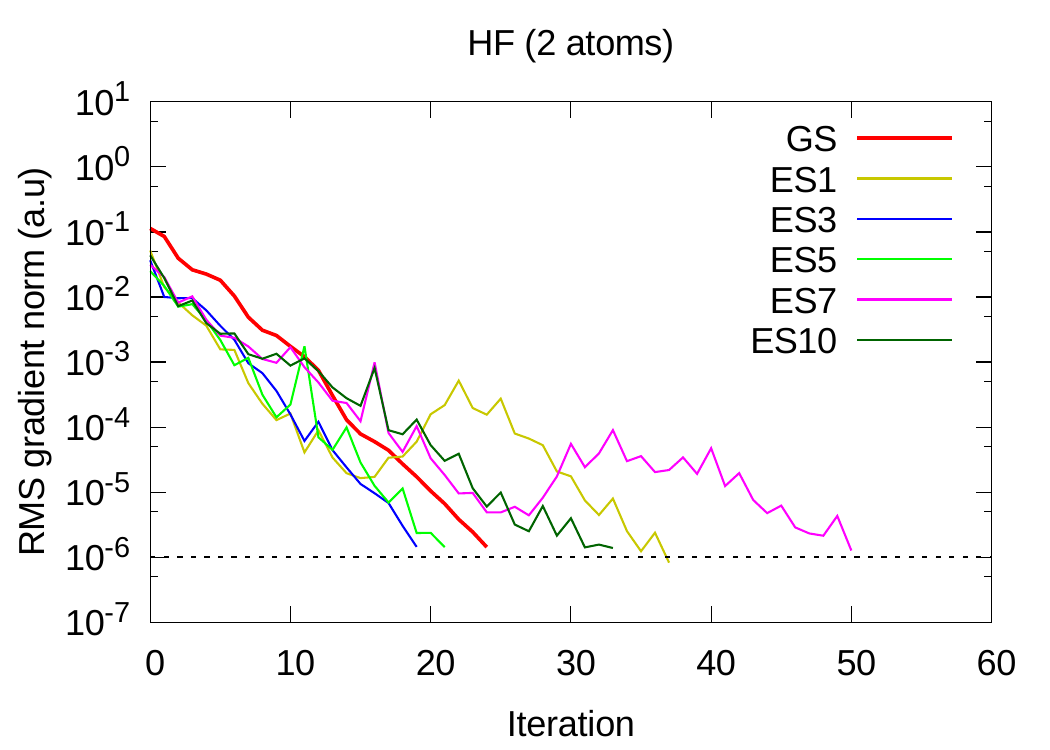}
    \includegraphics[width=0.30\textwidth]{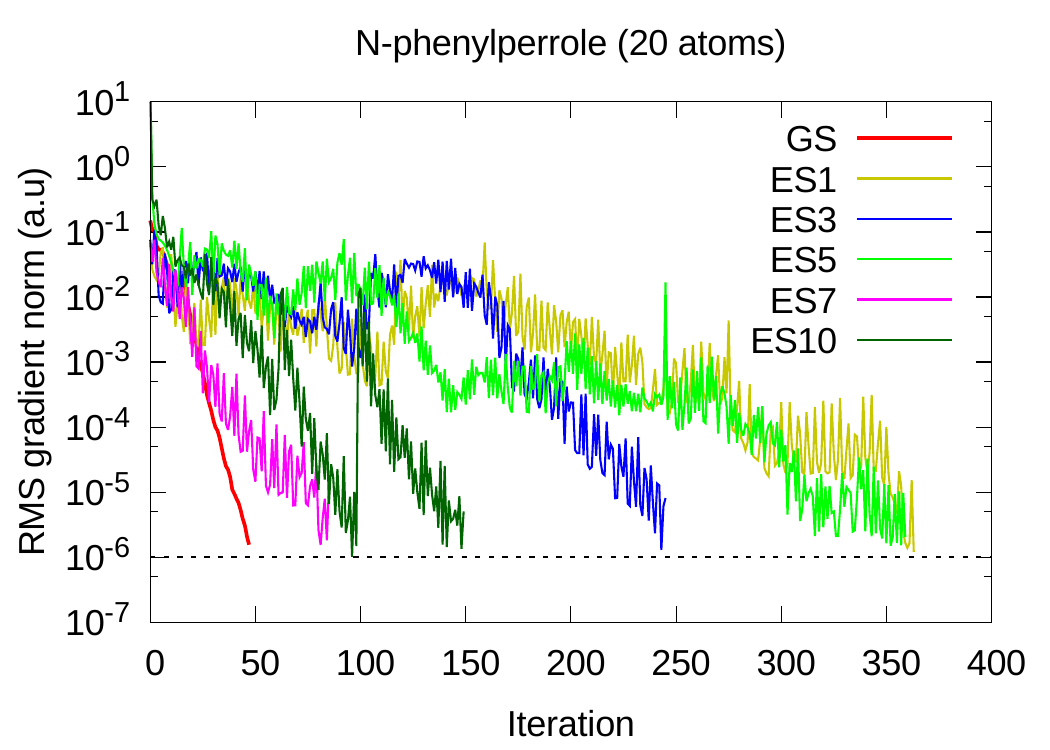}	
    \includegraphics[width=0.30\textwidth]{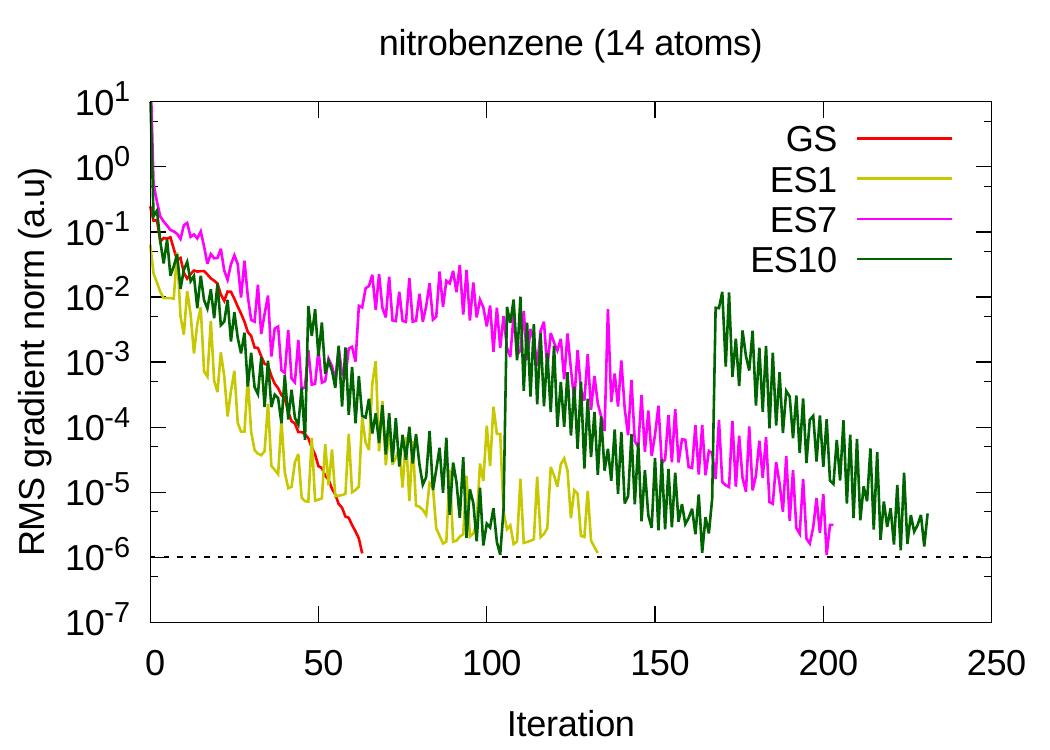}
    \includegraphics[width=0.30\textwidth]{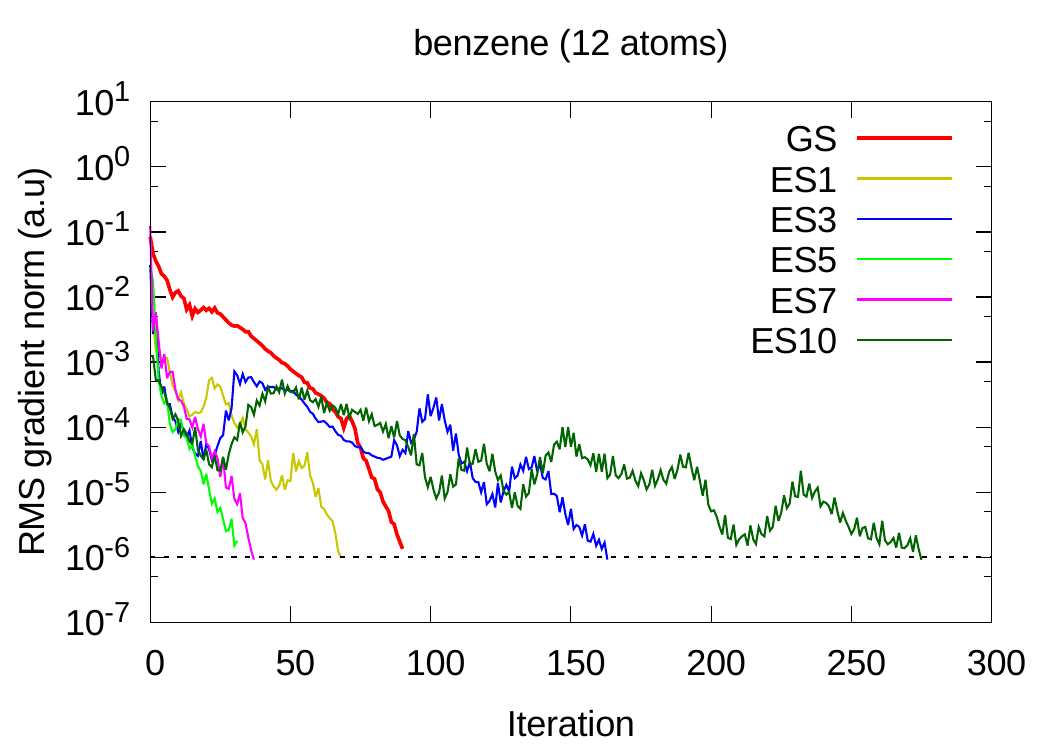}
    \includegraphics[width=0.30\textwidth]{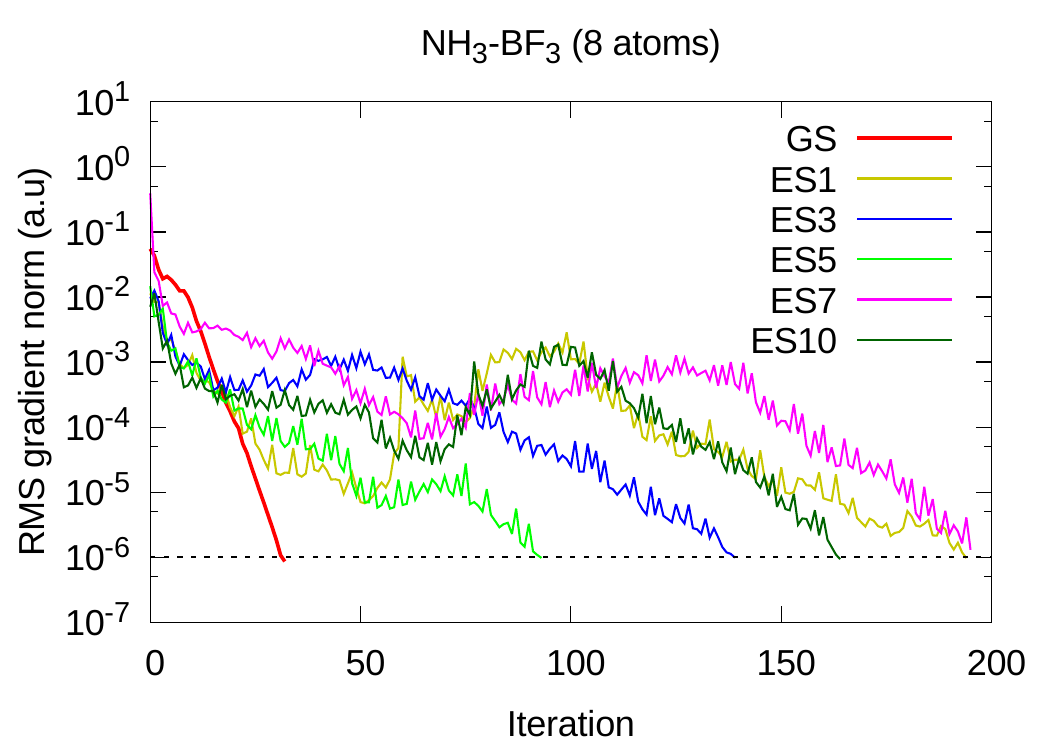}
  \caption{Convergence rate for the VM TIDFT procedure for the ground states (GS) and selected excited states (ES). Line search steps, one per PCG step, are not included in the iteration count. The dashed lines show the SCF convergence threshold.}
\label{fig:scf}
\end{figure*}

\label{subsec:convergence}
\textbf{SCF convergence.} Fig.~\ref{fig:scf} shows how the norm of the gradient of the loss functional changes during the PCG optimization of the ground and excited states. The default VM SCF optimization algorithm achieves stable convergence for the ground state~\cite{pham2024direct}. The algorithm also works well for the optimization of excited-state orbitals, sometimes converging faster than the optimization of the ground state. This is because an excited-state optimization already starts with the well-converged canonical orbitals of the ground state. For some systems, however, the optimization of the excited states takes many more iterations to converge as the optimizer needs to find the right balance between maintaining orthogonality to the previously optimized states and minimizing the energy. Converging excited-state optimization also takes longer when the penalty strength update is necessary to enforce stricter orthogonality between the electronic states. These updates, performed in an outer loop of the main PCG optimization, can be seen as sharp increases of the gradient norm from the convergence line in Fig.~\ref{fig:scf}.
%RZK after thesis submission: adjust C_P on every iteration, not in the outer loop.

It should be noted that a large number of iterations is required to achieve convergence even for the ground state of simple systems. This suggests that the rate of convergence can be improved by replacing PCG -- one of the simplest algorithm for unconstrained optimization -- with quasi-Newton or trust-region optimizers that are known to work well for the ground-state KS DFT~\cite{voorhis2002geometric}. Additionally, faster optimization can be achieved by updating the penalty strength on each iteration, instead of updating it in the outer loop.

Finally, it is worth mentioning that VM TIDFT method does not guarantee to find the lowest-energy excitations. As with all other orbital-optimized DFT methods~\cite{hait2021orbital}, the result of the optimization will be a stationary point lying in the basin of the initial guess.

\section{Conclusions}

VM TIDFT is a simple DFT method designed to prevent collapse of excited states during their variational optimization. The main idea of VM TIDFT is to allow nonorthogonal electronic states in the optimization process but gradually push them towards orthogonality with a simple continuous penalty function. With nonorthogonal orbitals and electronic states, VM TIDFT can use molecular orbital coefficients as independent variables in the optimization procedure. This in turn leads to simple closed-form analytical expressions for the gradient and allows to employ any of the ubiquitous unconstrained optimization algorithms that guarantees convergence of the excited-state optimization. Numerical tests on multiple molecular systems show that VM TIDFT optimization performed with the PCG algorithm is robust and the method computes accurate energies for well-behaved excitations and, unlike TDDFT, for more challenging CT and double-electron excitations.

Due to the simplicity of VM TIDFT, this approach can be implemented using other unconstrained minimization techniques, such as quasi-Newton or trust-region methods. This can improve the computational efficiency of VM DFT. VM TIDFT is also ideally suited to implement the full variational ROKS optimization of spin-purified open-shell singlet states~\cite{frank1998molecular, filatov1999density, pham2024direct}, which are necessary for more accurate description of many excited states. Additionally, unconstrained VM TIDFT optimization makes the computation of atomic forces in excited states straightforward, facilitating computation of emission spectra and dynamical photophysical and photochemical processes in excited states.

From a broader perspective, variable-metric optimization can be readily extended to electronic structure methods where the use of nonorthogonal wavefunctions is unavoidable (e.g. optimization of strictly localized orbitals~\cite{burger2008linear, khaliullin2013efficient, peng2013effective, shi2021robust}), or where linear dependencies present a serious problem during variational optimization (e.g. multi-configurational wavefunctions). 

\section{Acknowledgements}

We would like to thank Zhenzhe Zhang for providing optimized structures for our tests. The research was funded by the Natural Sciences and Engineering Research Council of Canada (NSERC) through Discovery Grant.
%RZK: Update grant number later (RGPIN-2016-0505). 
The authors are grateful to Compute Canada for computer resources allocated under the CFI John R. Evans Leaders Fund program.

%\bibliography{manuscript_ES} 
\providecommand{\latin}[1]{#1}
\makeatletter
\providecommand{\doi}
  {\begingroup\let\do\@makeother\dospecials
  \catcode`\{=1 \catcode`\}=2 \doi@aux}
\providecommand{\doi@aux}[1]{\endgroup\texttt{#1}}
\makeatother
\providecommand*\mcitethebibliography{\thebibliography}
\csname @ifundefined\endcsname{endmcitethebibliography}
  {\let\endmcitethebibliography\endthebibliography}{}

\end{document}